# Regulating the Hydrophobic Domain in Peptide-Catecholamine Coassembled Nanostructures for Fluorescence Enhancement

*Ruoyang Zhao,* [a, b, ‡] *Feng Gao,* [a, ‡] *Maoyu Li,* [c, ‡] *Xingkun Niu,* [c, ‡] *Shihao Liu,* [a] *Xinmin Zhao,* [c] *Liping Wang,* [a, *] *Jun Guo* [a, *] *and Feng Zhang,* [a, c, *]

[a] Wenzhou Institute, University of Chinese Academy of Sciences, Wenzhou 325001, China.

[b] College of Life Science, University of Chinese Academy of Sciences, Beijing 101408, China.

[c] Key Laboratory of Optical Technology and Instrument for Medicine, Ministry of Education, University of Shanghai for Science and Technology, Shanghai 200093, China.

[‡] These authors contributed equally.

ABSTRACT

Hydrophobic domains provide specific microenvironment for essential functional activities in life. Herein, we studied how the coassembling of peptides with catecholamines regulate the hydrophobic domain-containing nanostructures for fluorescence enhancement. By peptide encoding and coassembling with catecholamines of different hydrophilicities, a series of hierarchical assembling systems were constructed. In combination with molecular dynamics simulation, we experimentally discovered the hydrophobic domain of chromophore microenvironment regulates the fluorescence of coassembled nanostructures. Our results shed light on the rational design of fluorescent bio-coassembled nanoprobes for biomedical applications.

Water plays key roles in chemical reactions-based life activities [1], which might account for the theory that life originates from the ocean. However, the isolation from water produces hydrophobic domains which often provide local microenvironments for protein/peptide folding and extending the lifetime of excitons, otherwise it will cause misfolding-induced diseases [2-6] e.g., Alzheimer's disease, Parkinson's disease, and the low quantum yield of fluorescence [7-12]. Life tunes the distribution and arrangement of hydrophobic groups as a natural color palette for creating different pigments [13]. In natural proteins, many sequences are used to create hydrophobic microenvironments, e.g., fluorescent proteins normally form a typical hydrophobic β-barrel structure to increase the fluorescence quantum yield [14, 15]. Tyrosine-containing peptides were successfully engineered into fluorescent nanomaterials, yet the artificial fluorescence still needs improvement [16-20]. To this end, a feasible approach taught by nature could be constructing hydrophobic domains to improve the quantum yield, and creating larger conjugated systems to make more red-shift. In this way, we selected the structurally diverse natural catecholamines as chromophores. In principle, with the π-π conjugation between the aromatic rings of tyrosine and the cation-π interaction between the amino group of lysine and the aromatic ring of tyrosine, strong aggregation ability can be achieved. In addition to coassembling with the amphiphilic tripeptide glycine-tyrosine-lysine (GYK), a series of fluorescent nanostructures were constructed with good stability. The results showed that catecholamines carrying hydrophobic groups as chromophores can effectively isolate water from the coassembled core domain, thus enhance the fluorescence quantum yield, and simultaneously, the conjugated system has been enlarged and eventually resulted in an effective redshift of emission (**Scheme 1**).

**Scheme 1. Catecholamine-like molecules coassemble with short peptides to form fluorescent nanostructures with tunable optical properties.** The structural differences among different catecholamine-like molecules mainly lie in the hydrophobicity of the amino-terminal modification group (R). The size and enhanced hydrophobicity of the group help to isolate water molecules from the environment, thereby improving the fluorescence performance of coassemblies.

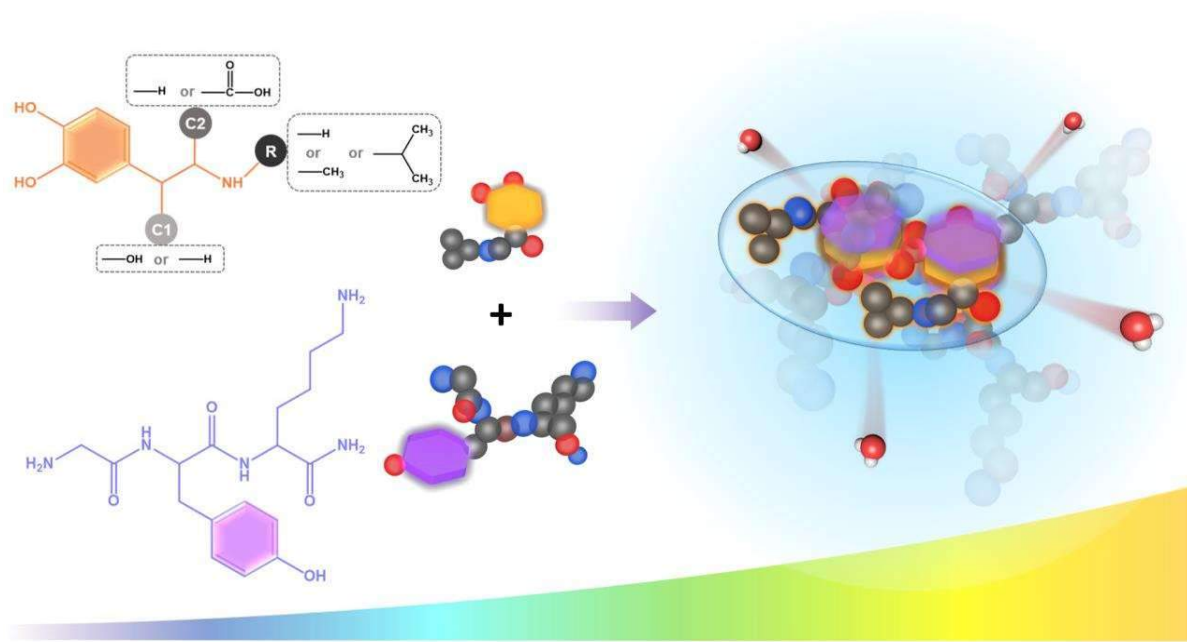

Catecholamines can easily oxidize and assemble into green fluorescent nanostructures in alkaline buffer solutions. However, generally the quantum yield of fluorescence is low, therefore increasing the hydrophobic groups could be an effective way for fluorescence enhancement (**Figure 1a**). Tyrosine and tyrosine-containing peptides have been used for constructing fluorescent nanomaterials [20]. However, their fluorescence performance primarily relies on the π-π interactions of the tyrosine side chain. Tyrosine hydroxylation forms DOPA, which, due to the increased phenolic hydroxyl groups, is prone to oxidative cross-linking [17]. In addition to the strong self-assembly tendencies, it is easier to form melanin-like structures [17, 18, 20]. DOPA often appears as a special residue in the sequence of biomimetic peptides, playing a role in polymerization and anchoring metal ions [21]. When coassembled with tripeptide GYK, it also promotes the formation of melanin-like aggregated structures, resulting in significant black precipitation and greatly reducing the fluorescence performance [17, 20]. Dopamine, formed through decarboxylation of DOPA, exhibits reduced assembly tendencies and can form biologically diverse materials, and as a typical catecholamine-like molecule, dopamine can provide important physicochemical properties such as antimicrobial, anti-inflammatory, antioxidant, and anticancer characteristics. They prefer to adhere to both organic and inorganic surfaces thus have been normally used in the material assembling [17]. In our previous research work, we synthesized bright green fluorescent nanostructures using dopamine and phenylalanine-containing short peptides [17]. However, in this study, the short peptides also contain tyrosine, and the phenolic hydroxyl groups to some extent increase the cross-linking with dopamine, resulting in decreased fluorescence intensity compared to assemblies with phenylalanine [17]. Similarly, the fluorescence performance of the assembly of norepinephrine and GYK did not improve. This demonstrated that hydroxylation at the C1 carbon contributes minimally to the coassemblies (**Figure 1b**). In contrast, the addition of N-methyl not only caused a redshift, but also resulted in a higher fluorescence intensity in the nanostructures coassembled by norepinephrine and the tripeptide GYK. As hydrophobic groups continued to increase, i.e., the fluorescence of coassemblies of the GYK with isoprenaline further redshifted and significantly enhanced the intensity. The acetylation of the amino group, in addition to enhancing hydrophobicity, also inhibited melanin synthesis pathways, weakening the cross-linking of catecholamine-like molecules, thereby only allowing the formation of quinones. From the molecular structure observed through HPLC, the introduction of hydrophobic groups forms a more compact assembled structure, which was crucial for increasing the π-π conjugation system (**Figure 1c**).

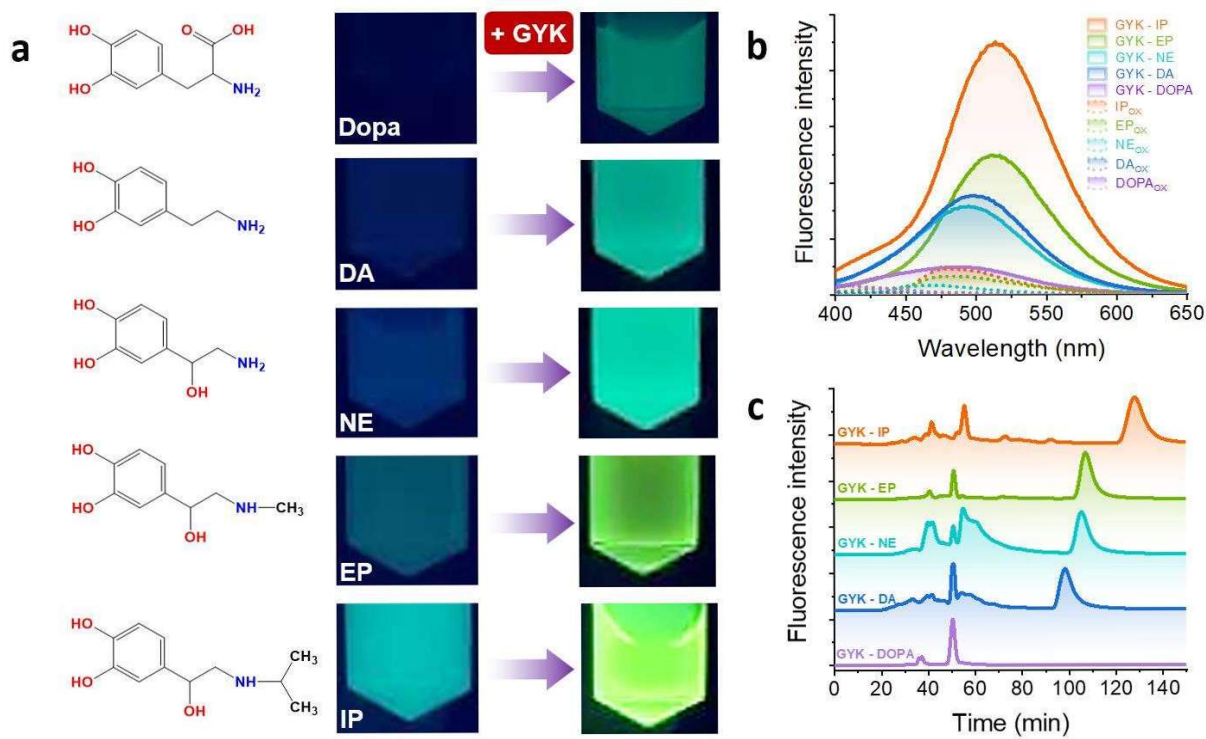

**Figure 1. Characterization of coassembled structures of different catecholamine-like molecules with short peptides.** a. Catecholamine-like molecules' structures and their UV-excited fluorescence photographs of coassembled structures with tripeptides. b. Fluorescence spectra of self-assembled structures of different catecholamine-like molecules upon oxidation and their coassembled structures with tripeptides. c. The size exclusion chromatography (SEC) assay of tripeptide GYK coassembled structures. The SEC was performed with a Superdex™ 30 Increase column (GE) on an HPLC system (Hitachi). DA, dopamine; Dopa, 3-(3,4-Dihydroxyphenyl)-DL-alanine; NE, norepinephrine; EP, epinephrine; IP, isoprenaline.

Non-covalent interactions for functional assemblies are of both diversity and reversibility, endowing supramolecular materials with diverse and controllable structures that can respond to external stimuli. Alkaline buffer conditions facilitate both the oxidation of phenolic hydroxyl groups and the deprotonation of amino groups, and coassembly can arise from the synergistic effect of these two processes. In this sense, we further investigated the fluorescence of three typical coassembled nanostructures (**Figure 2a**). Specifically, for coassembling with GYK, the amino group of dopamine not only helped enhance water solubility, but also supported a weak and flexible driving force due to its small size. The hydrophobic groups provided strong assembling tendencies, leading to the formation of larger conjugated systems and hydrophobic cavities to accommodate chromophores. Size distribution analysis using dynamic light scattering (DLS) showed that the coassembled nanostructures formed through oxidation were larger (**Figure 2b**). On the other hand, GYK possessed amphiphilic characteristics, with abundant amino or amide groups providing a hydrophilic shell for coassembling. As the amino group of tyrosine formed a peptide bond, pre-assembled peptides would not form melanin intermediates, and the oxidation of side-chain into quinones was restricted, all enabling good dispersibility of the coassembled nanostructures in solution. Regarding to the principle of aggregation induced emission (AIE), intramolecular mechanical motion will consume the excitation energy, leading to inefficient emission [7, 22]. Although oxidation crosslinking can limit molecular rotation to some extent and enable the formation of smaller-sized nanostructures in a short time, this way could active the melanin pathway and result in fluorescence quenching as the assembling time prolongs. However, under the aggregation state induced by acetylation of the amino group, the formation of hydrophobic cavities isolates the π-π conjugated system from the water environment, and the movement of aromatic groups was spatially restricted. In this way, the assemblies released the excitation energy through the radiation pathway, displaying increasing fluorescence emission (**Figure 2c**). Overall, the enhancement of emission performance is determined by the synergistic effect of assembly and oxidation.

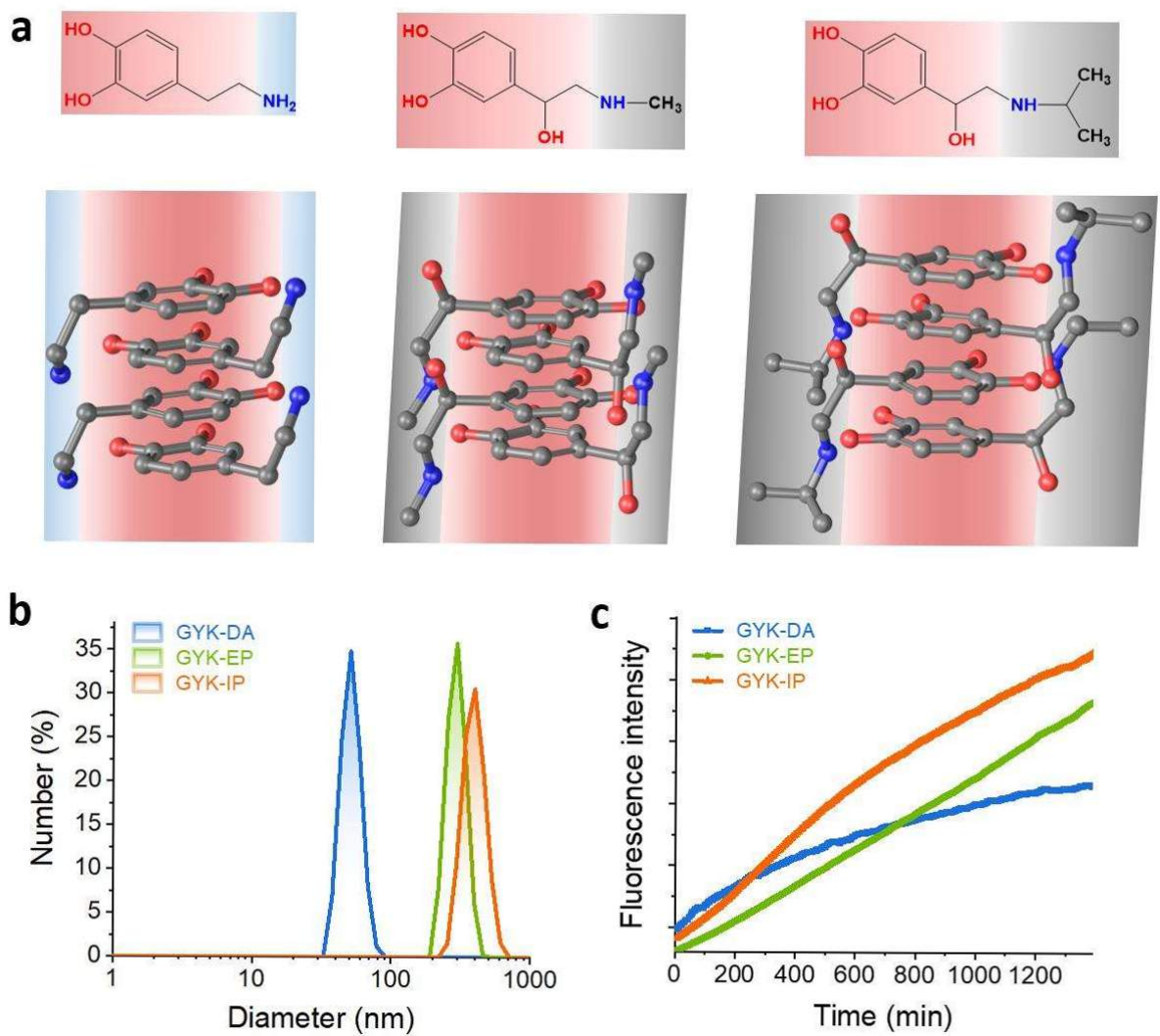

**Figure 2. The influence of hydrophobic cavities on coassembled structures.** a. Schematic representation of the core cavities of different catecholamine molecules and their coassemblies with tripeptides. The red color represents the π-π conjugated region, the blue color represents the hydrophilic region, and the gray color represents the hydrophobic region. b. Size analysis of the three coassembled structures. c. Assembling dynamics of the three fluorescent structures.

During the process of protein folding to form natural structures, a large number of assembly structures at different hierarchical levels will generate, ranging from enzymes to light-harvesting complexes to ionic channels, existing throughout the entire biological system. Inspired by nature, designing self-assembled structures that mimic the functionality of natural macromolecules can address certain challenges associated with biomaterials, such as improving their stability and recognition efficiency. Peptides with β-sheet characteristics are a common and easily accessible choice [1, 14, 23-30]. By modifying the β-barrel sequence, the fluorescence system can be effectively regulated [14, 15], enabling the construction of biosensors based on specific recognition. The coassembled fluorescent nanostructure of GYK-IP is formed under deprotonation conditions. A decrease in pH in the environment affects the stability of the coassembled nanostructure, leading to fluorescence quenching. Since the assembly was carried out under alkaline conditions, the coassembled nanostructures were mainly dependent on the action of non-covalent bonds, so the process was reversible. The experiment confirmed that no obvious crosslinked structures were found after the decomposition of each group with the decrease of pH (Figure S1-S7). To enhance the fluorescence performance, we designed a simpler β-sheet structure inspired by the β-barrel of fluorescent proteins [14, 15]. First, we used positively charged lysine residues as the hydrophilic region, while consecutive isoleucine residues served as the hydrophobic region, creating an amphiphilic molecule. Second, to establish the growth site of the assembled structure, we introduced tyrosine residues into the sequence. Third, to minimize the steric hindrance, we introduced glycine residues with maximum conformational flexibility between the tyrosine residue and the hydrophobic region. Finally, to further enhance the assembling propensity, both N-terminal acetylation and C-terminal amidation were added to the peptide sequence, Ac-IIIGYK-NH$_2$. The β-sheet hydrogen bonding between the main chains of the short peptides drives the aggregation in the y-axis direction (**Figure 3a**). The side chain residues in the x-axis direction provide strong hydrophobic interactions. Due to the electrostatic repulsion of the lysine residues, the most likely arrangement between adjacent molecules should be antiparallel β-sheet, which was confirmed by molecular dynamics simulations (**Figure 3b**, Figure S8). CD spectra also confirmed that the structure of the short peptide tends to be β-sheet (Figure S9)。

Lysine residues were distributed freely on the surface of the coassembled nanostructure, while the remaining five residues formed an antiparallel β-sheet structure. The theoretical length of this portion is 1.71 nm, slightly smaller than the total length (2.16 nm) of the hexapeptide (**Figure 3c**). Atomic force microscopy (AFM) imaging results showed that the IIIGYK peptide formed ribbon-like rigid nanofibers with a height of approximately 1.5 nm and a width of about 100 nm (**Figure 3d-e**), consistent with the theoretical values. The small difference could be attributed to dried samples, which confirmed that the short peptides was standing upright on the mica surface along the z-direction in the way of template-assisted self-assembly [1, 8, 28-30], and the fibers extended and grown in the y-direction driven by hydrogen bonding in the β-sheet structure (**Figure 3f**). The hydrophobic interaction facilitates the extension in the x-direction, forming wide nanoribbons. Due to the abundant positively charged lysine residues distributed on the fiber surface, there was a strong affinity for the negatively charged surface of mica, and even the growth of nanofibers aligned with the lattice direction of mica can be observed. On the other hand, due to charge repulsion between fibers, the growth along the z-axis is inhibited, and we did not observe the formation of thick fibers (larger diameter). Instead, multiple fibers are arranged parallel to the lattice direction (**Figure 3d**).

After successfully achieving the desired one-dimensional nanoscale structures, we introduced GYK-IP for forming higher-order hierarchical coassembled structures [6, 29]. AFM imaging results showed that the nanofibers grown with discrete heights on the existing structure, increasing by approximately 2 nm per layer. Moreover, they no longer strictly aligned with the lattice direction of mica (**Figure 3g-h**). This indicated the growth of structures with a thickness of 0.4-0.5 nm between the self-assembled hexapeptide nanoribbons. Specifically, GYK-IP formed new types of coassembled nanostructures with the tyrosine residue. Due to the nanoribbon constraint, fluorescence nanostructures larger than hundreds of nanometers are no longer formed (**Figure 2b**). Instead, GYK-IP participated in the layer-by-layer assembled nanofibers, forming new hybrid fluorescent nanostructures (**Figure 3i**). The hydrophobic β-sheet layer stabilized the coassembled structures, enhances their fluorescence, all made it suitable for biomedical applications to resist pH changing and fluorescence quenching (Figure S10).

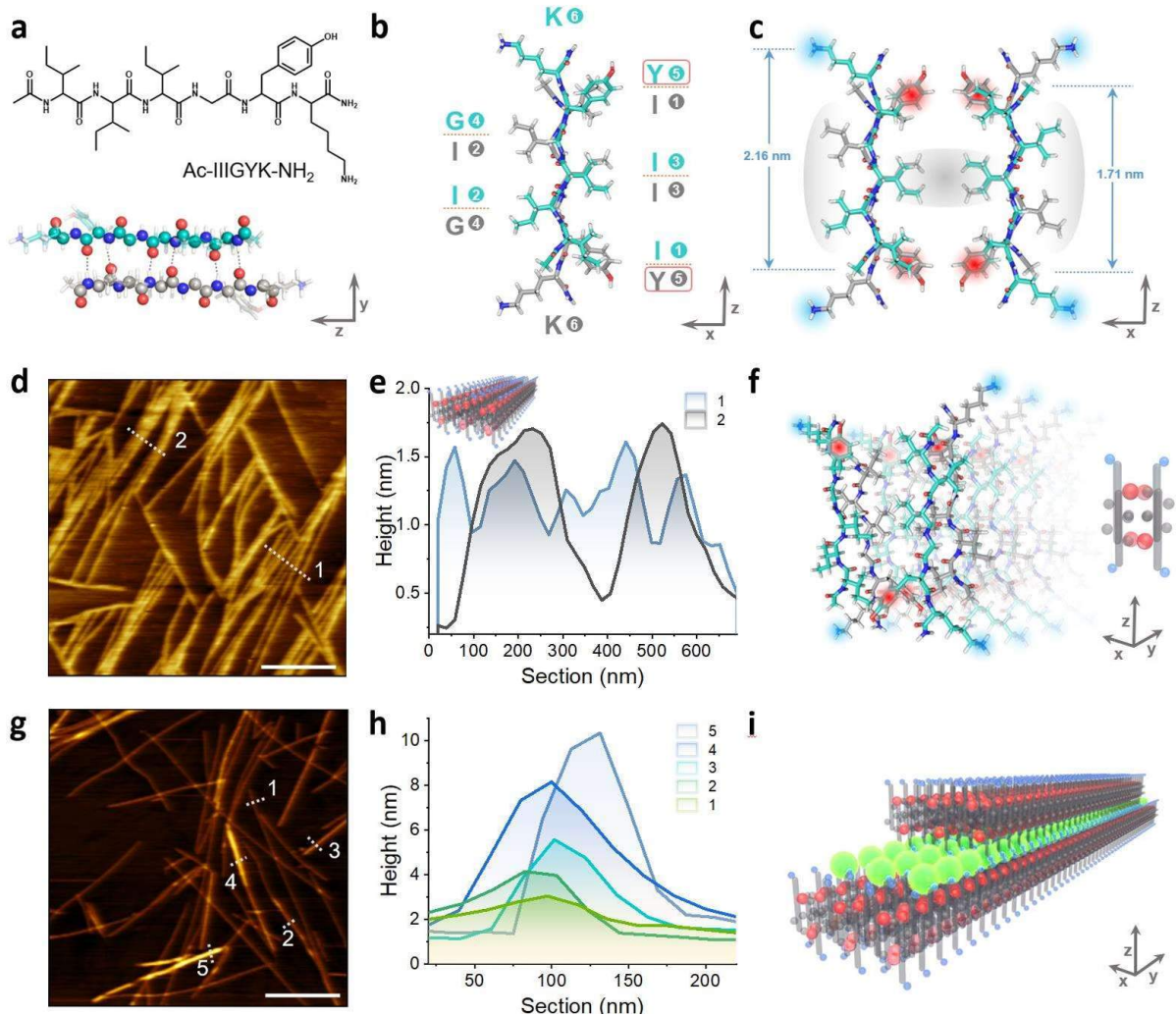

**Figure 3. Hierarchical assembled structures of hexapeptides.** a. Molecular structure of Ac-IIIGYK-NH$_2$ and schematic representation of hydrogen bonding between main chains of the hexapeptides. b-c. Schematic representation of antiparallel aggregation between hexapeptides, where carbon atoms are shown in cyan or gray, oxygen atoms in red, nitrogen atoms in blue, and hydrogen atoms in white. d-e. AFM height image of nanofibers formed by the hexapeptide Ac-IIIGYK-NH$_2$, along with the corresponding height distribution along the fiber axis. Scale bar, 1 μm. f. Schematic representation of the hexapeptide standing upright on the mica surface. g-h. AFM height image of nanofibers formed by the hexapeptide Ac-IIIGYK-NH$_2$ coassembling with GYK-IP, along with the corresponding height distribution along the fiber axis. Scale bar, 1 μm.

i. Schematic representation of the three types of coassembled structures, where red-blue-gray represents the simplified Ac-IIIGYK-NH$_2$ molecule, and the green fluorescent spheres represent the coassembled fluorescent nanostructures of GYK-IP.

The coassembled system of peptide nanofibers and fluorescent nanoparticles exhibits high flexibility and stability, making it suitable for a wide range of applications[19, 31-33]. For instance, the nanofiber surface is rich in positively charged amino groups, which endows it with excellent transmembrane potential. Additionally, tyrosine residues are highly responsive to tyrosinase and can rapidly crosslink with each other. Melanoma cells typically express high levels of tyrosinase, so this characteristic can be utilized to modulate the cellular activity of melanoma [20]. We conducted tests in B16 melanoma cells and used tyrosinase-deficient lung cancer cells (A549) and normal cells (L929) as controls. The results showed that the coassembled nanostructures exhibited specific cytotoxic effects on melanoma cells, suggesting their potential use in anticancer drug development (**Figure 4**, Figure S11).

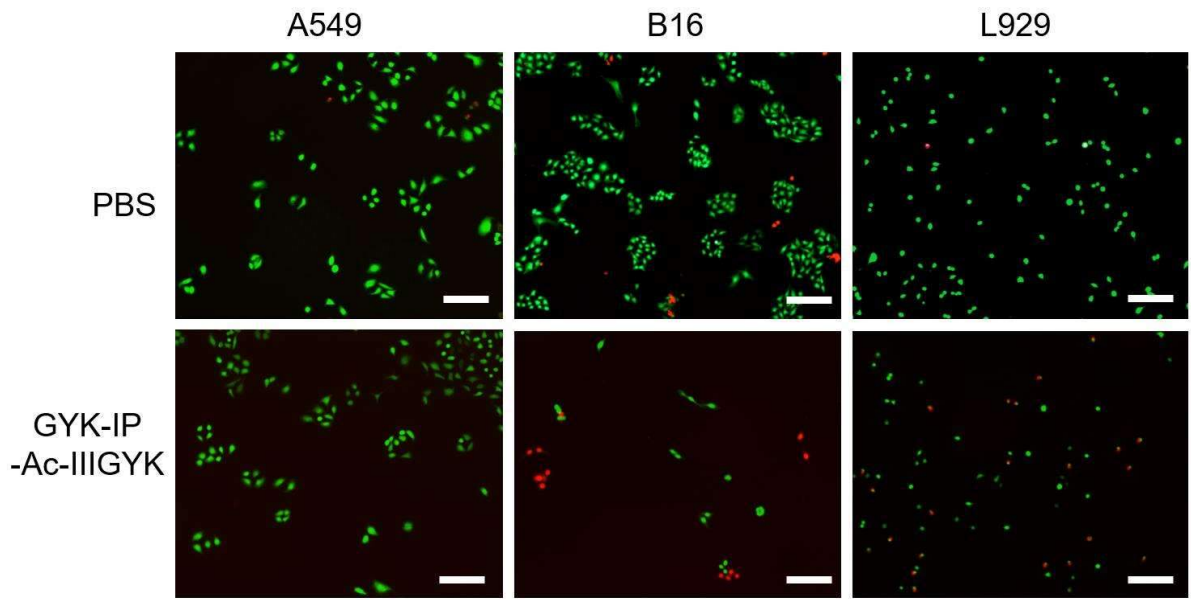

**Figure 4. Peptide coassembled nanofibers affect the cell growth.** Live cells are shown in green, while dead cells are shown in red. All scale bar are 0.2 mm.

By designing short peptides suitable for hierarchical assembly, we have demonstrated the crucial regulatory role of hydrophobic moieties in fluorescent coassembled nanostructures. The hydrophobic groups within the molecule self-assemble to functional domains, and the addition of hydrophobic moieties from catecholamines protects the phenolic hydroxyl groups and inhibits cross-linking, leading to the formation of the first-level assemblies. The coassembly involving tyrosine-containing amphiphilic tripeptides further strengthens the conjugated system, enhancing the optical properties of chromophores and resulting in a significant redshift, forming the second-level assemblies. In the third-level assemblies, we utilize the rational distribution of continuous hydrophobic side chains to construct antiparallel β-sheet structures as protein-inspired shells. This enables the encapsulation of fluorescent nanomaterials and the formation of larger composite structures. The synergistic effects of different peptide sequences allow for the orderly layer-by-layer assembly of multiple β-sheet layers, resembling the synthesis of native proteins. Such multilevel structures can serve as universal templates for designing artificial proteins and functional biomolecular arrays, facilitating the construction of more complex coassembled hybrid nanostructures.

ASSOCIATED CONTENT

**Supporting Information**.

AUTHOR INFORMATION

**Corresponding Author**

\***Feng Zhang** − Wenzhou Institute, University of Chinese Academy of Sciences, Wenzhou 325001, China; Quantum Biophotonic Lab, Key Laboratory of Optical Technology and Instrument for Medicine, Ministry of Education, School of Optical-Electrical and Computer Engineering, University of Shanghai for Science and Technology, Shanghai 200093, China; orcid.org/0000-0001-6035-4829; Email: fzhang@usst.edu.cn

\***Jun Guo** − Wenzhou Institute, University of Chinese Academy of Sciences, Wenzhou 325001, China; orcid.org/0000-0001-6944-0731; Email: guojun-nbm@wiucas.ac.cn

\***Liping Wang** − Wenzhou Institute, University of Chinese Academy of Sciences, Wenzhou 325001, China; orcid.org/0000-0002-9282-2751; Email: lpwangwiucas@ucas.ac.cn

**Authors**

**Ruoyang Zhao** – Wenzhou Institute, University of Chinese Academy of Sciences, Wenzhou 325001, China; College of Life Science, University of Chinese Academy of Sciences, Beijing 101408, China.

Feng Gao – Wenzhou Institute, University of Chinese Academy of Sciences, Wenzhou 325001, China.

**Maoyu Li** – Quantum Biophotonic Lab, Key Laboratory of Optical Technology and Instrument for Medicine, Ministry of Education, School of Optical-Electrical and Computer Engineering, University of Shanghai for Science and Technology, Shanghai 200093, China.

**Xingkun Niu** – Quantum Biophotonic Lab, Key Laboratory of Optical Technology and Instrument for Medicine, Ministry of Education, School of Optical-Electrical and Computer Engineering, University of Shanghai for Science and Technology, Shanghai 200093, China.

**Shihao Liu –** Wenzhou Institute, University of Chinese Academy of Sciences, Wenzhou 325001, China.

**Xinmin Zhao –** Quantum Biophotonic Lab, Key Laboratory of Optical Technology and Instrument for Medicine, Ministry of Education, School of Optical-Electrical and Computer Engineering, University of Shanghai for Science and Technology, Shanghai 200093, China.

‡ R.Z., F.G., M.L. and X.N. contributed equally to this work. The manuscript was written through contributions of all authors. All authors have given approval to the final version of the manuscript.

ACKNOWLEDGMENT

This work was supported by the National Key Research and Development Program of China (No. 2021YFA1200400), the National Natural Science Foundation of China (No. 32271298 and T2241002), and Wenzhou Institute, University of Chinese Academy of Sciences (WIUCASQD2021003).

ABBREVIATIONS

DA, Dopamine; Dopa, 3-(3,4-Dihydroxyphenyl)-DL-alanine; NE, Norepinephrine; EP, Epinephrine; IP, Isoprenaline.

**Table of Contents**

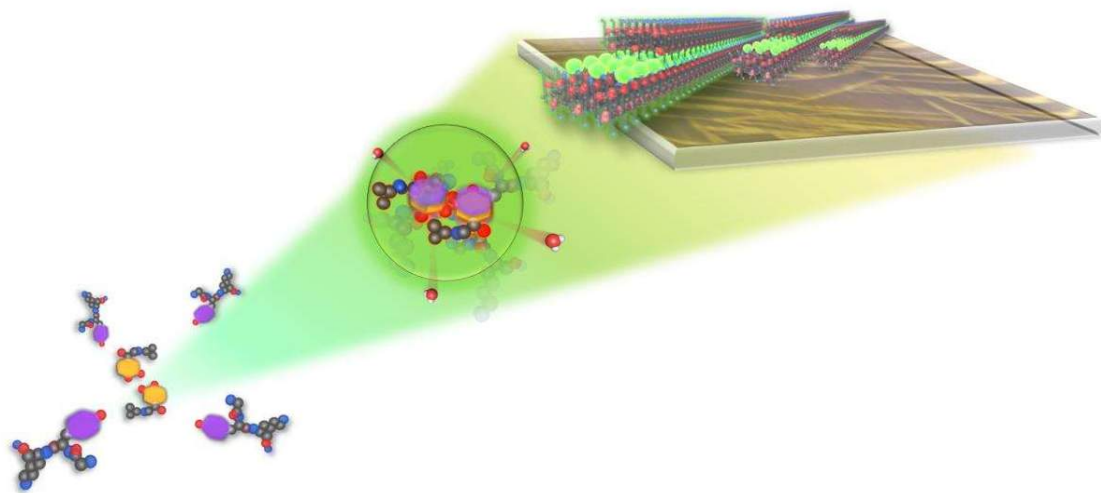